\documentclass[twocolumn,showpacs,%
  nofootinbib,aps,superscriptaddress,%
  prd,notitlepage,showkeys,10pt]{revtex4-1}

\usepackage{amssymb}
\usepackage{amsmath}
\usepackage{graphicx}
\usepackage{dcolumn}
\usepackage{hyperref}

\begin{document}

\title{Tripartite Counterfactual Quantum Cryptography}
\author{Hatim Salih}
\email[]{salih.hatim@gmail.com}
\affiliation{Qubet Research, London NW6 1RE, UK}

\date{\today}

\begin{abstract}
We show how two distrustful parties, ``Bob'' and ``Charlie'', can share a secret key with the help of a mutually trusted ``Alice'', counterfactually---that is with no information-carrying particles travelling between any of the three parties.  
\end{abstract}


\maketitle

In a recent paper \cite{Shenoy}, a quantum cryptography protocol was proposed where an entrusted Alice allows Bob, a bank for example, and Charlie, a client unsure of Bob's identity, to share a secret key that not even Alice has access to. The protocol's aim of extending the original N09 counterfactual quantum key distribution (QKD) protocol \cite{Noh} to three parties is theoretically and practically interesting. But even though no photons travel all the way between Bob and Charlie, making the protocol counterfactual in one sense, an eavesdropper, Eve, still has full access to Alice's information carrying photons, making the protocol not counterfactual in another crucial sense. This does not in itself make the protocol unsecured, but the powerful promise of security \cite{Noh} based on the total absence of information-carrying photons from the transmission channels is lost. 

Here, we show using a two-cycle chained quantum Zeno effect (CQZE) \cite{Salih}, how Alice can enable Bob and Charlie to share a secret key with no information-carrying photons traveling between any of the three parties---achieving complete counterfactuality. Security arguments \cite{Noh} and proofs \cite{Yin} based on complete counterfactuality should thus hold. 

\begin{figure}
\centering
\includegraphics[width=0.5\textwidth]{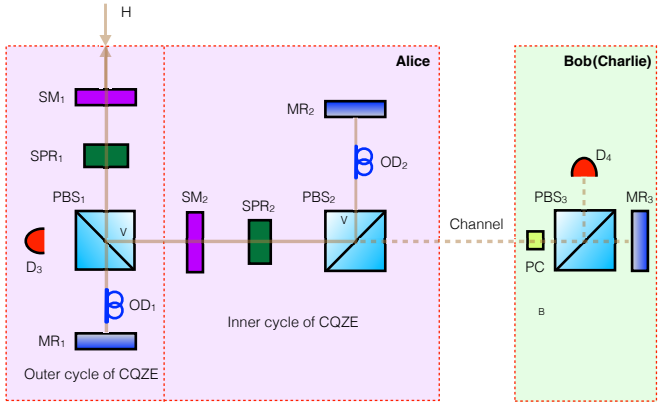}
\caption{\label{fig: TriOne}The chained quantum Zeno effect (CQZE). Bob(Charlie) can block the channel by switching Pockels cell $PC$ on, directing the photon towards detector $D_4$. Initially, switchable mirror $SM_1$ is switched off allowing Alice's $H$ photon in before being switched on again. Switchable polarisation rotator $SPR_1$ then applies the following rotation to the photon, $\left| \text{H} \right\rangle \to 1/\sqrt2(\left| \text{H} \right\rangle + \left| \text{V} \right\rangle)$, before being switched off for the rest of this outer cycle. Polarising beamsplitter $PBS_1$ reflects the $V$ part of the superposition towards Bob(Charlie). (Optical delays $OD$ ensure that the effective path lengths correctly match.) Switchable mirror $SM_2$ is then switched off to allow the $V$ part of the superposition into the inner interferometer before being switched on again. Switchable polarisation rotator $SPR_2$ then applies the following rotation, $\left| \text{V} \right\rangle \to 1/\sqrt2(\left| \text{V} \right\rangle - \left| \text{H} \right\rangle)$, before being switched off for the rest of this inner cycle. Polarising beamsplitter $PBS_2$ reflects the $V$ part of the superposition while passing the $H$ part towards Bob(Charlie). There are now two scenarios: (i) If Bob(Charlie) blocks the channel, effectively making a measurement, the part of the photon superposition inside the inner interferometer ends up in the state $\left| \text{V} \right\rangle$, unless the photon is lost to $D_4$. The same applies to the next inner cycle. Switchable mirror $SM_2$ is then switched off to allow this part of the superposition, whose state has remained $\left| \text{V} \right\rangle$, out. In the next outer cycle, $SPR_1$ rotates the photon's polarisation from $1/\sqrt2(\left| \text{H} \right\rangle + \left| \text{V} \right\rangle)$ all the way to $\left| \text{V} \right\rangle$ before being switched off for the rest of this outer cycle. $PBS_1$ reflects the photon towards Bob(Charlie). As before, after two inner cycles, the photon remains in the state $\left| \text{V} \right\rangle$ unless it is lost to $D_4$. $SM_1$ is then switched off to allow the photon, whose final state is now $\left| \text{V} \right\rangle$, out. (ii) If instead Bob(Charlie) does not block the channel, the part of the photon superposition in the inner interferometer, namely $1/\sqrt2(\left| \text{V} \right\rangle - \left| \text{H} \right\rangle)$,  will be rotated all the way to the sate $-\left| \text{H} \right\rangle$ after two inner cycles. Switchable mirror $SM_2$ is then switched off to allow this part of the superposition out. Measurement by $D_3$ leaves the photon in the overall state $\left| \text{H} \right\rangle$ moving towards $SM_1$, unless it is lost to $D_3$. The same applies to the next outer cycle (and two inner cycles). $SM_1$ is then switched off to allow the photon, whose final state is $\left| \text{H} \right\rangle$, out. Counterfactuality is ensured as any photon going into the channel would either trigger $D_3$ for the case of Bob(Charlie) not blocking, or else trigger $D_4$ for the case of Bob(Charlie) blocking.}
\end{figure}

The overall action of the two-cycle CQZE, whose inner working is explained in the caption of FIG. \ref{fig: TriOne}, on Alice's horizontally ($H$) polarised photon is the following, $\left| \text{H} \right\rangle \to \left| \text{H} \right\rangle$ when the channel is not blocked, and $\left| \text{H} \right\rangle \to \left| \text{V} \right\rangle$ when the channel is blocked. Crucially, in both cases the photon does not travel through the channel. We know this because a photon going into the channel would either trigger detector $D_4$, for the case of Bob(Charlie) blocking, or else trigger detector $D_3$ for the case of Bob(Charlie) not blocking. Note that with the smallest possible number of cycles used (two inner and two outer cycles) the probability of the photon not being lost due to detection by $D_3$ or $D_4$ is $\approx$ $1/5$ \cite{Probability}. (This can be made arbitrarily close to one by increasing the number of cycles, but at the cost of practicality.)

\begin{figure}
\centering
\includegraphics[width=0.5\textwidth]{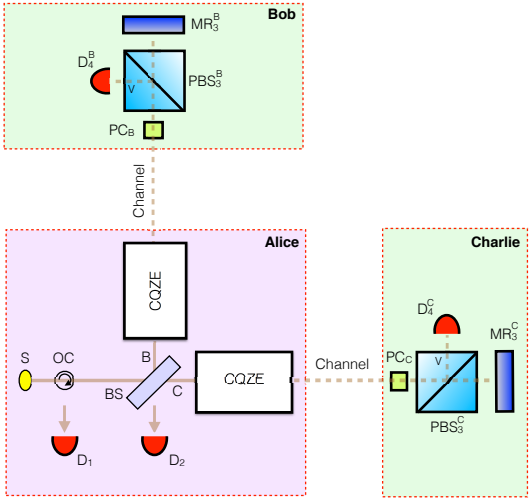}
\caption{\label{fig: Protocol}Protocol for tripartite counterfactual quantum cryptography. Bob(Charlie) randomly encodes a ``0''(``1'') by not blocking his channel and a ``1''(``0'') by blocking it. Bob(Charlie) can block his channel by switching Pockels cell $PC_{B(C)}$ on, which flips polarisation, directing the photon towards $D_4$. Initially, Alice sends a $H$ photon from her photon source $S$ towards beamsplitter $BS$, which puts the photon in an equal superposition of being on path B (leading to Bob) and path C (leading to Charlie). If Bob and Charlie encode different bit values, the two parts of the photon superposition reflected back towards Alice's $BS$ (from top and from right) will be, by the action of the two CQZEs, identically polarised. Constructive interference therefore takes place resulting in Alice's $D_2$ clicking with certainty (provided the photon was not lost to $D_3$ or $D_4$). If, however, Bob and Charlie encode the same bit value, the two parts of the photon superposition reflected back towards Alice's $BS$ will be oppositely polarised. Interference does not take place. $D_1$ and $D_2$ are therefore equally likely to click. A click at $D_1$ uniquely corresponds to Bob and Charlie randomly agreeing in their bit choices. (Here, $OC$ stands for optical circulator, which directs a photon exiting left towards $D_1$.)}
\end{figure}

\emph{Protocol for tripartite counterfactual quantum cryptography}---Alice starts by sending a $H$ photon from the left towards beamsplitter $BS$ of the Michelson interferometer of FIG. \ref{fig: Protocol}, which applies a $\pi/2$ rotation to the path qubit, putting the photon in an equal superposition of being on path B (leading to Bob) and path C (leading to Charlie). Bob(Charlie) encodes a ``0''(``1'') by not blocking his channel and encodes a ``1''(``0'') by blocking it. If they encode different bit values, the two parts of the photon superposition reflected back towards Alice's $BS$ (from top and from right) will be, by the action of the two CQZEs, identically polarised. Constructive interference therefore takes place resulting in Alice's $D_2$ clicking with certainty (provided the photon was not lost to $D_3$ or $D_4$). If, however, Bob and Charlie encode the same bit value, the two parts of the photon superposition reflected back towards Alice's $BS$ will be oppositely polarised. Interference does not take place because differing polarisation acts as a which-path ``tag''. $D_1$ and $D_2$ are therefore equally likely to click. Since $D_1$ clicking corresponds uniquely to Bob and Charlie randomly agreeing on their bit value, whenever $D_1$ clicks Alice publicly instructs Bob and Charlie to keep the corresponding bits as their sifted key, the rest are discarded. Throughout, no information-carrying photons have traversed either channel.

In summary, using a two-cycle chained quantum Zeno effect, we have shown how to achieve completely counterfactual QKD between two distrustful parties assisted by another entrusted party---with no information-carrying particles travelling between any of them.

%\begin{equation}
%\begin{split}
%\label{Exiting BS}
%&1/\sqrt2(\lambda \left| \nwarrow \right\rangle + \mu \left| \swarrow \right\rangle) \otimes \left| \text{lower path} \right\rangle + \\
%&1/\sqrt2(\lambda \left| \nwarrow \right\rangle - \mu \left| \swarrow \right\rangle) \otimes \left| \text{upper path} \right\rangle.
%\end{split}
%\end{equation}
     
\begin{acknowledgments}
This work is partially supported by Qubet Research, a start-up in quantum information.
\end{acknowledgments}

%\clearpage

\begin{thebibliography}{99}
\bibitem{Probability} The probability that the photon avoids detection by $D_3$ for the case of Bob not blocking is given by ${\cos}^{2M}{\theta}_{M}$, where $M$ is the number of outer cycles. For $M=2$ we get a probability of $1/4$.  On the other hand, the probability that the photon avoids detection by $D_4$ for the case of Bob blocking is given by $\prod_{m=1}^{M} (1-{\sin}^{2}{m\theta}_{M}{\sin}^{2}{\theta}_{N})^N$, where $M(N)$ is the number of outer(inner) cycles. For $M=2$ and $N=2$ we get a probability of $9/64$. The overall probability of the photon making it back to Alice is therefore $25/128$.
\end{thebibliography}
\end{document}